\documentclass[aps,pre,onecolumn,superscriptaddress,longbibliography,10pt]{revtex4-2}

\usepackage{hyperref}
\usepackage{nameref}
\usepackage{amsmath,amssymb}
\usepackage{graphicx}
\usepackage{dcolumn}
\usepackage{bm}
\usepackage[dvipsnames]{xcolor}
\usepackage{media9}

\graphicspath{ {Figures/}}

\begin{document}

\title{Interacting particles crossing barriers: spatial statistics, choice of route taken and role of interactions}

\author{Gulzar Ahmad}
\affiliation{Department of Mathematical Sciences and Interdisciplinary Centre for Mathematical Modelling, Loughborough University, Loughborough, LE11 3TU, United Kingdom}
\affiliation{Higher Education Department, Govt.~of the Punjab, Lahore, 54000, Pakistan}
\author{Andrew J.~Archer}
\affiliation{Department of Mathematical Sciences and Interdisciplinary Centre for Mathematical Modelling, Loughborough University, Loughborough, LE11 3TU, United Kingdom}
\author{Sergey Savel’ev}
\affiliation{%
 Department of Physics, Loughborough University, Loughborough, LE11 3TU, United Kingdom}
\author{Marco G.~Mazza}
\affiliation{Department of Mathematical Sciences and Interdisciplinary Centre for Mathematical Modelling, Loughborough University, Loughborough, LE11 3TU, United Kingdom}

\date{\today}

\begin{abstract}
We consider the statistics of pairs of particles moving in a potential energy landscape exhibiting a choice of different pathways from a local minimum start point to the global energy minimum destination.
We examine the impact of both attractive and repulsive interactions between the particles, treating these via either a soft Gaussian or a Lennard-Jones potential. We investigate the statistics of which particle arrives at the destination first, arrival-time distributions and where in the system the second particle is located when the first arrives. We find that the typical arrival time of the first particle can be significantly different to that of a single particle, sometimes being much longer and in other cases being shorter, depending on the nature of the particle interactions.
In the case of soft Gaussian attractive interactions, which effectively bind the pair of particles together to form a single larger particle, the arrival time can be much longer, since bigger particles cross barriers more slowly. 
In contrast, the arrival time is shorter when there is a repulsive core interaction between the particles, due to excluded-area and particle-ordering effects. 
We investigate the probability distribution of particles' positions through the Jensen--Shannon divergence which unveils the dependence of the spatial distribution of the particles within the confining potential on the parameters of the particle interaction potentials.
\end{abstract}

\maketitle
\section{Introduction}

Escaping from a potential well and crossing a barrier is a remarkably important problem in physics and chemistry. It underpins the dynamics of atoms and molecules and processes ranging from Brownian particles in a potential well to enzyme catalysis to Josephson junctions and more \cite{kramers1940brownian,hanggi1990reaction,mel1991kramers}.

While most studies focus on the dynamics of a single particle trapped in a potential well, it is also interesting to study the effect of interactions when multiple atoms or molecules are present and simultaneously attempting to escape the well, driven by thermal fluctuations. The dynamics of permanently bound particles (dimers) in a deterministic, anharmonic potential were investigated \cite{fugmann2008deterministic}, and in the presence of noise \cite{asfaw2012exploring}, where a non-monotonic dependence of the escape rate on the binding constant was observed.
Brownian motors (ratchets) have also been studied in the context of interparticle interactions, which, depending on the coupling, can even lead to a reversal of probability currents \cite{reimann2002brownian,menche2006two}. 
The problem of long polymers confined in the metastable side of a double-well potential has also been addressed \cite{sebastian2000kramers}. 
More recently, Singh studied coupled particles in one-dimensional potentials \cite{singh2017kramers,singh2017escape}. 

In this work, we are interested in slightly more realistic potential energy landscapes, that can often be complex and rugged, featuring numerous saddle points, barriers of different heights, and multiple paths connecting stable states \cite{cammarota2015spontaneous, fitzgerald2023stochastic}, which arise in condensed matter systems, such as glassy systems, colloidal fluids, or gels. At the most basic level, a complex potential energy landscape exhibits competing pathways and metastable minima.

Specifically, we investigate how a pair of particles explore and cross the barriers in the energy landscape previously studied in Refs.~\cite{fitzgerald2023stochastic, ahmad2025nonequilibrium}, where the dynamics of a single particle was investigated. We give all the necessary details of the potential below, but the key features of the potential are that is has a choice of two routes for the particles to reach the global minimum in the potential. One route is longer, but goes over a lower barrier (i.e.\ lower saddle point), while the other route is shorter, but is over a higher barrier.
One of the main findings of Refs.~\cite{fitzgerald2023stochastic, ahmad2025nonequilibrium} is that if the time available for barrier crossing is short, then if barrier crossing is to occur, the most likely route from one minimum to the other is the path over the higher barrier. In contrast, if the time available is longer, then the most likely route is over the lower barrier.
Here, we consider a pair of particles confined within this potential and how the particle interactions influence which particle arrives in the destination minimum first and the spatial statistics of the second particle, at the moment when the first arrives.
We find that the nature of the particle interactions greatly influences these quantities, with order of magnitude changes in arrival times possible, depending on the nature of the interactions.

We investigate here the effect of an interaction potential $V(r)$ between two Brownian particles separated by a distance $r$ and navigating a background potential energy landscape $\phi(\bm{x})$. We compare two very different interparticle potentials $V(r)$.
First, we consider the soft Gaussian core model $V_\mathrm{G}(r) \propto \exp[{-{r^2}/{\lambda_\mathrm{G}^2}}]$, with range parameter $\lambda_\mathrm{G}$, originally introduced by Stillinger in the context of crystallization of charged colloidal particles or polymer solutions \cite{stillinger1976phase}, and is now widely used for complex fluids \cite{louis2000mean, likos2001effective}. Second, we consider a Lennard-Jones pair potential $V_\mathrm{LJ}(r)$.

In recent years, transition path times have garnered considerable attention \cite{hummer2004transition,berezhkovskii2005one,Dudko2006,sega2007,zhang2007transition,chung2009experimental,chaudhury2010harmonic,orland2011generating,kim2015mean,makarov2015shapes,Truex2015,daldrop2016transition,pollak2016transition,Neupane2016,caraglio2018influence,Neupane2018} due to their ability to offer valuable information on structural transformations, such as those occurring during protein folding and phase transitions~\cite{cossio2018transition}. However, the majority of studies have predominantly focused on one-dimensional energy landscapes or two-dimensional systems involving a single particle, leaving the exploration of more complex, multiparticle systems largely under-explored.

The rest of the paper is organized as follows. Section~\ref{sec:2} describes the two simple interaction potentials used in this work, and the equations governing the dynamics of our system. In Sec.~\ref{sec:results} we present results for spatial distributions about particle positions, for the two particles, under varying settings, and discussion on time taken for the particle to reach the destination. Finally, Sec.~\ref{sec:6}, concludes the paper with some final remarks and proposed further work.

\section{Model}
\label{sec:2}

\subsection{Potential energy landscape}

\begin{figure}
    \centering
    \includegraphics[width=0.5\linewidth]{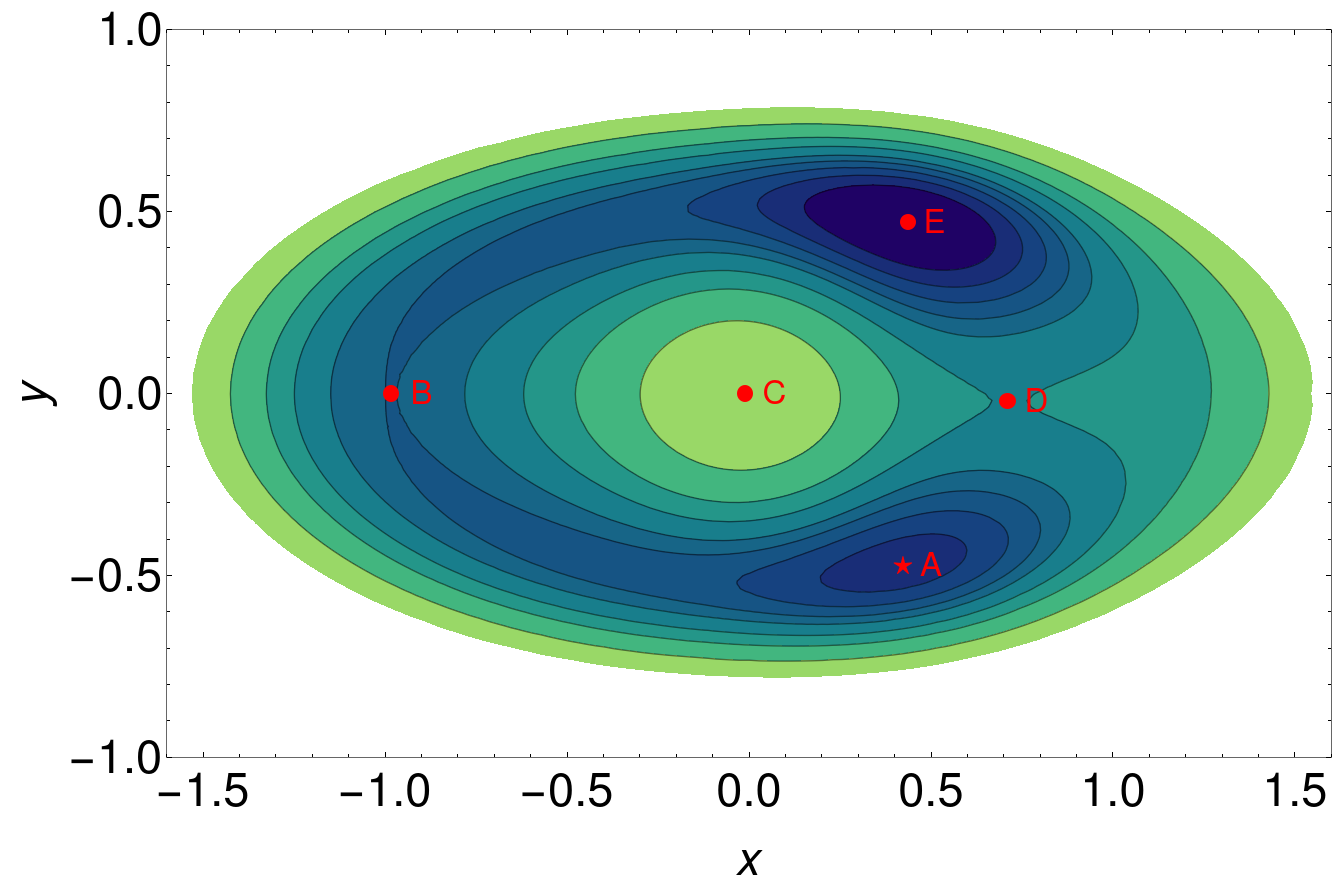}
    \caption{Contour plot of the potential $\phi(x, y)$ in Eq.~\eqref{eq:potential}, highlighting its stationary points -- see Eq.~\eqref{eq:points}.}
    \label{fig:potential}
\end{figure}

We study a system of two particles moving in two dimensions with coordinates $\bm{x}=(x,y)$, that experience a complex potential energy landscape~\cite{ahmad2025nonequilibrium}, displayed in Fig.~\ref{fig:potential}.
The dimensionless potential is defined as  
\begin{equation}\label{eq:potential}
  \beta_\mathrm{ref} \phi(x,y)=\beta_\mathrm{ref}\phi_\mathrm{M}(x,y)-\frac{1}{2} \frac{x}{\ell}+\beta_\mathrm{ref}\phi_G(x,y),
\end{equation}
where  $\beta_\mathrm{ref}\equiv (k_\mathrm{B} T_\mathrm{ref})^{-1}$, $T_\mathrm{ref}$ is a reference temperature,  $k_\mathrm{B}$ Boltzmann's constant, and $\ell$ is a reference length-scale in our model.
In the following, we use a dimensionless temperature $T^*=T/T_\mathrm{ref}$ and set the unit of length $\ell=1$.
The first term in Eq.~\eqref{eq:potential} is a stretched Mexican-hat type potential
\begin{equation}
    \beta_\mathrm{ref}\phi_\mathrm{M}(x,y)=4 \left(x^{2}+4 y^{2}-1\right)^{2}\,.
\end{equation}
The second term generates a small constant force parallel to the $x$-axis, and the third term
\begin{align} \label{eq:pot}
 \beta_\mathrm{ref} \phi_G(x,y) = & -2 {\mathrm e}^{-4 \left(x -\frac{1}{2}\right)^{2}-4 \left(y -\frac{1}{2}\right)^{2}} 
-{\mathrm e}^{-4 \left(x -\frac{1}{2}\right)^{2}-4 \left(y +\frac{1}{2}\right)^{2}} 
+3{\mathrm e}^{-4 \left(x -1\right)^{2}-4 y^{2}}+4 {\mathrm e}^{-10 \left(x^{2}+y^{2}\right)}
\end{align}
consists of four Gaussian contributions to generate local maxima and minima in the potential at specific points. The final term in $\phi_G$ is added to make the local maximum near the origin even higher and thus suppress the probability of the system passing near this point $C$, and instead more often taking either of the two paths near the saddle points $B$ and $D$; see Fig.~\ref{fig:potential}.

The potential energy $\phi(x,y)$ exhibits five stationary points with coordinates
\begin{align}
\label{eq:points}
\begin{split}
    A&=(x_A, y_A)=(0.413, -0.471)\,, \\
    B&=(x_B, y_B)=(-0.984, -0.000)\,, \\
    C&=(x_C, y_C)=(-0.017, 0.003)\,, \\
    D&=(x_D, y_D)=(0.710, -0.022)\,, \\
    E&=(x_E, y_E)=(0.436, 0.460).
\end{split}
\end{align}
The point $A$ is a local minimum, while $E$ is the global minimum. The point $C$ is a local maximum, and the points $B$ and $D$ are saddle points.

Given these features of the potential in Eq.~\eqref{eq:potential}, a particle moving under the influence of $\phi(x,y)$ and starting in the potential well around point $A$ will typically follow one of two pathways from $A$ to $E$.
The first pathway involves passing in the vicinity of the saddle point $D$, whereas the second pathway goes through the vicinity of point $B$, the other saddle point.
Passing near $B$ requires the system to travel a longer distance, resulting in a longer pathway time.

A particle starting from point $A$  and going through the saddle at point $D$ has to overcome a barrier height $\beta_\mathrm{ref}\Delta \phi_{AD}\equiv \beta_\mathrm{ref}(\phi(x_D,y_D)-\phi(x_A,y_A))=2.7$. In contrast, the barrier height for going up to the saddle at point $B$ from initial point $A$ is $\beta_\mathrm{ref}\Delta \phi_{AB}\equiv \beta_\mathrm{ref}(\phi(x_B,y_B)-\phi(x_A,y_A))=1.3$.
Thus, the system has two choices: either (\textit{i}) go over  the higher saddle at point $D$, with a relatively short distance to travel in state space, or (\textit{ii}) go over the lower saddle point $B$, which however  involves a longer journey.

\subsection{Interparticle potential}

To investigate the role of interactions between pairs of Brownian particles in the barrier crossing dynamics, we consider two paradigmatic interparticle potentials. The first is the Gaussian potential
\begin{align} \label{eq:gpp}
    V_\mathrm{G}(r)=\varepsilon_\mathrm{G} e^{-{r^2}/{\lambda_\mathrm{G}^2}} ,
\end{align}
where the parameter $\lambda_G$ determines the range, $\varepsilon_G$ determines the interaction strength, and $r\equiv |\bm{x}_1-\bm{x}_2|$, where $\bm{x}_1$ and $\bm{x}_2$ are the positions of the two Brownian particles.
When $\varepsilon_G>0$, this potential is often referred to as the Gaussian core model and it provides a good approximation for the effective interaction between the centres of mass of polymers in solution, under good solvent conditions~\cite{likos2001effective}. 
In this case, $\lambda_G$ corresponds to the polymer's radius of gyration, $\varepsilon_G$ is an energy scale on the order of $k_BT$~\cite{likos2001effective} representing the strength of the interparticle interaction.  
Here we consider both the case $\varepsilon_G>0$, corresponding to repulsive interactions, and also the case $\varepsilon_G<0$, corresponding to an attractive interaction.

We also investigate the Lennard-Jones (LJ) potential
\begin{equation} \label{eq:LJ_GF}
    V_\mathrm{LJ}(r)=4\varepsilon_\mathrm{LJ}\left[ \left(\frac{\sigma_\mathrm{LJ}}{r}\right)^{12}-\left(\frac{\sigma_\mathrm{LJ}}{r}\right)^6\right]\,,
\end{equation}
where $\varepsilon_\mathrm{LJ}$ is the Lennard-Jones energy scale (depth of the potential well), and $\sigma_\mathrm{LJ}$ the particle diameter. The potential minimum is at $r_{eq}=2^{\frac{1}{6}}\sigma_{LJ}$.

\subsection{Stochastic dynamics}

We study a system of two Brownian particles governed by the following overdamped Langevin equations
\begin{align} \label{eq:OvDIP}
     \gamma \frac{\partial\boldsymbol{x}_k}{\partial t} = -\frac{\partial }{\partial \boldsymbol{x}_k}\phi(\boldsymbol{x}_k) -\frac{\partial }{\partial \boldsymbol{x}_k}V(|\boldsymbol{x}_1-\boldsymbol{x}_2|) + \boldsymbol{\xi}_k(t),
\end{align} 
where $k=\{1,2\}$, $\gamma$ is the friction coefficient and the random force $\boldsymbol{\xi}_k$ represent independent Gaussian white noise.
Henceforth, $\gamma$ is set equal to one, i.e.\ is absorbed into the timescale.
Thus, the natural Brownian timescale for the system $\tau_B=\beta_{\rm ref}\gamma\ell^2$, is also set equal to one.
The random components of $\boldsymbol{\xi}(t)$ (we dropped the subscript $k$ here deliberately) obey
\begin{align} \label{eq:EA}
\begin{split}
        \langle \xi_i (t) \rangle &=0 \,, \\
 \langle \xi_i (t) \xi_j(t') \rangle &=2\gamma k_\mathrm{B} T \delta_{ij}\delta(t-t')\,,
\end{split} 
\end{align}
where $\delta(t-t')$ is a Dirac delta distribution, $\delta_{ij}$ is the Kronecker delta, with $i,j =\{x,y\}$, and $\langle \cdot \rangle$ denotes the ensemble average.
The second term on the right-hand side of Eq.~\eqref{eq:OvDIP} is the force stemming from the interparticle potential. In the following, this term will be set equal to either $V_\mathrm{G}(r)$ or $V_\mathrm{LJ}(r)$ -- see below Secs.~\ref{sec:gauss} and \ref{sec:LJ}, respectively. 

Equations~\eqref{eq:OvDIP} and \eqref{eq:EA} are based on the assumption of a Markovian process, which means that the system has no memory. However, this is not always a trivial matter. The interested reader can consult the Markovianity test along a reaction coordinate proposed in~\cite{berezhkovskii2018single}.

\subsection{Simulation initialization and methods}
\label{sec:IID}

We initialize the system with both particles at or in proximity of point $A$ (a local minimum of $\phi(x,y)$) and then we numerically integrate Eqs.~\eqref{eq:OvDIP} and \eqref{eq:EA} using the standard Euler--Maruyama finite difference method \cite{kloeden1992stochastic}, 
with a time step $\Delta t=10^{-3}$, for a total of $2\times 10^7$ time steps. to increase statistical accuracy we performed $5\times 10^5$ independent runs. Spatial derivatives are computed on a square grid, with spacing $\Delta x=\Delta y = 0.01$.
In order to compute the first passage time to reach the global minimum at $E$, we define `reaching $E$' as the first time when a particle $k$ satisfies the condition $|\boldsymbol{x}_k-\boldsymbol{x}_E|<0.1$.

For the case when the particles interact via the Gaussian potential $V_\mathrm{G}(r)$, we set the initial position of both particles equal to point $A\equiv (x_A,y_A)$.
The Gaussian potential in Eq.~\eqref{eq:gpp} does not diverge as the particle separation $r\to0$, unlike typical interatomic or colloidal potentials, allowing configurations in which the centers of the two particles overlap. When this potential is used to model polymers in solution, this represents the centers of mass of two polymer chains overlapping, reflecting the  non-compact nature of polymers.
Also, even with overlapping centers of mass, chains can fluctuate without intersecting due to their soft, interpenetrable structure~\cite{likos2001effective}.
The sets of parameters for the Gaussian potential~\eqref{eq:gpp} explored in this work are $\varepsilon_G=\{-10,-5,0,5\}$ and $\lambda_G=\{0.2,0.5,0.8\}$. We thus explore both attractive and repulsive interparticle potentials.

The LJ potential diverges $V_{LJ}(r)\to\infty$ as $r \to 0$. Therefore, to avoid the singularity that arises from the overlapping of particles, we initialize the particles with a separation of $r_{eq}$ (potential minimum distance) around point $(x_A,y_A)$.
That is, we assign the initial positions as $(x_A+0.5r_{eq},y_A)$ and $(x_A-0.5r_{eq},y_A)$ and label these two points as $P_1$ and $P_2$, respectively.
Likewise, in this LJ case we refer to the particle starting at $P_1$ as `particle 1' and the other as `particle 2'.
During the simulation, to avoid excessively large forces that arise at small interaction distances $r$, causing numerical instability, we replace the interaction force for $r<r_c$, where $r_c=0.9\sigma_{LJ}$, with the value of the repulsive force at $r=r_c$. The value of $r_c$ is chosen to strike the right balance between avoiding singularity and preserving the dynamics of the pair potential and is consistent with best practices in molecular dynamics simulations, where similar short-range cut-offs are used to ensure numerical stability~\cite{allen2017computer}.
The sets of parameters for the LJ potential~\eqref{eq:LJ_GF} that we consider are: $\varepsilon_{LJ}=\{0,3.33,6.67,10\}$ and $\sigma_{LJ}=\{0.33,0.67,1.0\}$.
 
To study the features of spatial and temporal probability distributions of the two interacting particles, we considered both the Jensen--Shannon divergence (JSD)~\cite{endres2003new} and order statistics for reaching the minimum at $E$.
The JSD $J(p_1(\boldsymbol{x}) \!\parallel \! p_2(\boldsymbol{x}))$ is a symmetrized and smoothed version of the Kullback--Leibler divergence  $D(p_1(\boldsymbol{x}) \!\parallel\! p_2(\boldsymbol{x}))$ \cite{kullback1951information} and is defined as
\begin{align}
\begin{split}
    J(p_1(\boldsymbol{x})\!\parallel \! p_2(\boldsymbol{x}))=\frac{1}{2}D(p_1(\boldsymbol{x})\!\parallel \! M(\boldsymbol{x}))
     +\frac{1}{2}D(p_2(\boldsymbol{x})\!\parallel \! M(\boldsymbol{x})),
\end{split}\label{eq:JSD}
\end{align}
where $M(\boldsymbol{x})=\frac{1}{2}(p_1(\boldsymbol{x})+p_2(\boldsymbol{x}))$ is a mixture distribution of $p_1(\boldsymbol{x})$ and $p_2(\boldsymbol{x})$ and
\begin{equation}
D(p_1(\boldsymbol{x})\parallel p_2(\boldsymbol{x}))=\int p_1(\boldsymbol{x})\log\frac{p_1(\boldsymbol{x})}{p_2(\boldsymbol{x})}d\boldsymbol{x}.
\end{equation}
A value of JSD close to zero indicates a high similarity in the two distributions, while the larger the value, the more different the distributions.

We investigate the distribution of the arrival time $t$ for either of the two particles to be the first to reach the destination.
This problem is closely related to the first-passage time problem~\cite{redner2023first}.
The minimum of individual arrival times is an ordered statistics~\cite{abonongo2024exponentiated}
\begin{align}
    t_{(1)}=\mathrm{min}(t_1,t_2).
\end{align}
If the arrival times of each of the particles are independent events (which is the case when the interaction parameters $\varepsilon_{G}=0$ or $\varepsilon_{LJ}=0$, for Gaussian or LJ interactions, respectively), then the probability density function (PDF) of $t_{(1)}$ reads
\begin{align} \label{eq:ord_GF}
    p_{t_{(1)}}(t)=2p_1(t)[1-G_1(t)],
\end{align}
where $p_1(t)$ is the PDF of the arrival time for a single particle, and $G_1(t)$ is the cumulative distribution function (CDF) of $p_1(t)$. 
As a result, the arrival time distribution for pairs of noninteracting particles is shifted toward earlier times, with a higher concentration of probability mass in the lower $t$ range. In other words, $p_{t_{(1)}}(t)$, which is henceforth denoted as $p(t)$, is expected to be narrower than the arrival time distribution for a single particle, {\em assuming the particles are noninteracting.}

For example, if the arrival time for a single particle starting at point $A$ follows an exponential distribution, with decay parameter $\tau$, defined by
\begin{align} \label{eq:exp_d}
    p_1(t)=\frac{1}{\tau}e^{-\frac{t}{\tau}} 
\end{align}
for all $t >0$, $\tau >0$, this then gives $G_1(t)=1-e^{-\frac{t}{\tau}}$, and so Eq.~\eqref{eq:ord_GF} gives
\begin{equation} \label{eq:ord}
\begin{split}
    p_{t_{(1)}}(t)=&2\frac{1}{\tau}e^{-\frac{t}{\tau}} \left[1-\left(1-e^{-\frac{t}{\tau}}\right) \right] \\
    =&\left({\frac{\tau}{2}}\right)^{-1}e^{-{t}/{\frac{\tau}{2}}}.
\end{split}
\end{equation}
Note that in Ref.~\cite{ahmad2025nonequilibrium} it was demonstrated that, except at very early times, the expression in Eq.~\eqref{eq:exp_d} provides a good approximation for the single-particle arrival time distribution.
Equation \eqref{eq:ord} therefore implies that the probability density for the arrival time of the first of a pair of particles to arrive at the destination $E$ should exhibit a faster decay over time than the single-particle case.
As we show below, for interacting particles, this is not always the case.

\section{Results and discussion}
\label{sec:results}

This section presents our results for the dynamics defined in the previous section.
We discuss first the Gaussian pair interaction case, followed by the LJ potential case. 

\subsection{Gaussian pair potential}
\label{sec:gauss}

\subsubsection{Location}

\begin{figure*}[t!] 
    \centering
    \includegraphics[width=0.99\textwidth]{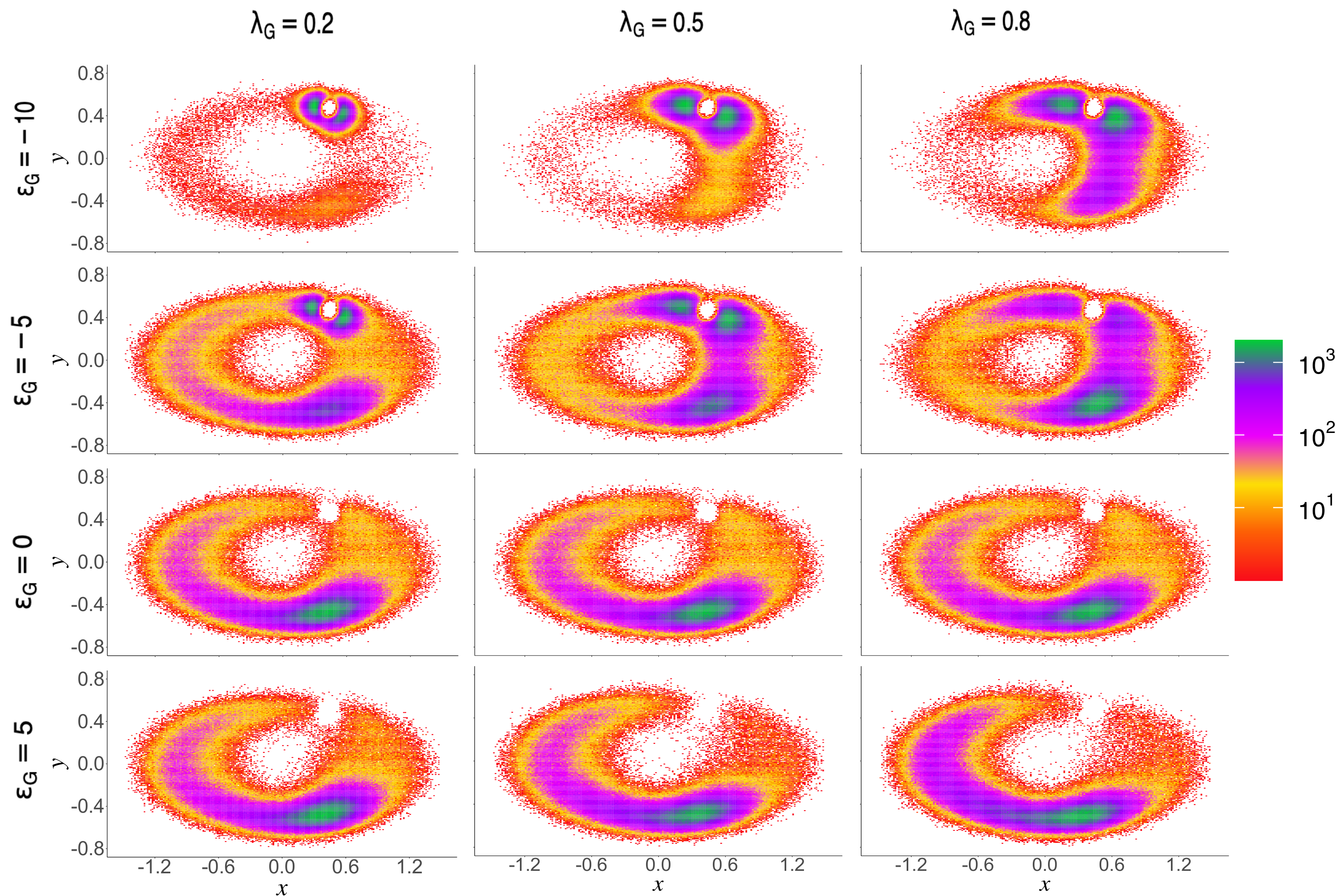}
    \caption{Probability distributions for the position of one particle at the moment in time when the other has reached the destination point $E$, for different values of interaction strength $\varepsilon_G$ and interaction range $\lambda_G$ of the Gaussian pair interaction potential defined in Eq.~\eqref{eq:gpp}. the results are from averaging over $5\times10^5$ independent runs.} 
    \label{fig:com_plot_G}
\end{figure*}

Figure~\ref{fig:com_plot_G} shows the two-dimensional distribution of the positions of one particle when the other has reached the destination. Operationally, reaching the destination point $E$ is defined as the time when the particle $k$'s distance from $E$ satisfies $|\boldsymbol{x}_k-\boldsymbol{x}_E|<0.1$. We observe that the position distribution of the particle not at point $E$ is highly sensitive to changes in the values of the pair interaction parameters, namely $\varepsilon_G$, the interaction strength, and $\lambda_G$, the interaction range, i.e.\ the `size' of the particles. The values of these define the dynamics of particles interacting through the Gaussian potential \eqref{eq:gpp}.

In the strongly attractive regime, e.g.\ the $\varepsilon_G=-10$ cases in Fig.~\ref{fig:com_plot_G}, we see that the second particle is predominantly localized in the region close to the destination point $E$, at the time when the first particle reaches the destination. This is because the strong attractive interaction has effectively `glued' the particles together, so that they predominantly move as a bound pair.
The density distribution of the second particle exhibits two prominent local maxima, one either side of point $E$, and a much smaller local maximum at point $A$, the start point, corresponding to the rare cases when one of the particles gets left behind at the start point by the other particle.

For the smallest value of the interaction range that we explored $\lambda_G=0.2$, the second particle's distribution is very tightly localized around the destination point $E$ when $\varepsilon_G<0$, sticking close to the first particle due to the strong attraction. 
Upon increasing $\lambda_G$, we observe that the distribution becomes broader around the destination point $E$. 
This can be attributed to the fact that although the pair of particles are tightly bound to one another, a larger interaction range (i.e.\ particle size) means that the distance between the pair of particles can be much greater.

In moderately attractive cases, such as the $\varepsilon_G=-5$ case in Fig.~\ref{fig:com_plot_G}, the second particle can detach from the first particle and remain more often in the vicinity of the start position (the local minimum $A$), while the first particle crosses over one of the barriers to point $E$. As a result, the distributions in Fig.~\ref{fig:com_plot_G} exhibits several prominent local maxima: one close to the start point $A$ and then a pair either side of the destination point $E$.
We can infer from this that the weaker attractive force allows the second particle to maintain greater independence, and its motion is less impacted by the dynamics of first particle.

In the absence of interaction, $\varepsilon_G=0$, Fig.~\ref{fig:com_plot_G} shows that the second particle behaves completely independently of the first and exhibits a very broad distribution, but mainly remains confined to the vicinity of its initial position $A$, regardless of the fact that the first particle has moved to the destination at the point $E$.
This behavior is expected because the lack of interaction.
Note that further details of the time evolution in the single particle case can be found in Refs.~\cite{fitzgerald2023stochastic, ahmad2025nonequilibrium}.

When the Gaussian interaction between the particles is repulsive, such as the $\varepsilon_G=5$ case, displayed in bottom row of Fig.~\ref{fig:com_plot_G}, we see that the distribution of the second particle is largely independent of the position of the first, with a broad distribution similar to the non-interacting case, except for in the case of large particle sizes $\lambda_G$.
In this repulsive and large interaction range scenario, 
we see an increased density around the saddle point $B$,
defining the barrier with a lower height at a longer distance.
This suggests that the preferred path of the second particle to reach the destination is through the saddle point $B$.
We note that the time to reach the destination for large particles is very small [see Figure~\ref{fig:t_e_s}(a)].
This finding suggests that repulsive forces push the particles apart such that the first particle to reach the destination at point $E$ effectively prevents the second particle from approaching the destination, particularly when the particle size is large.

These observations illustrate that when one particle reaches the destination first, the spatial distribution of the second particle changes in a systematic way from its initial position $A$ to the destination point $E$. As a result, strong repulsive forces drive separation, while attractive forces encourage particle cohesion, with the effect becoming more pronounced with an increase in particle size. 

In order to check the stability of the numerical method used for the time integration of Eq.~\eqref{eq:OvDIP}, we also produced some results with the smaller time step of $\Delta t = 10^{-4}$, which is an order of magnitude smaller than the value used to produce Fig.~\ref{fig:com_plot_G}. This smaller value of $\Delta t$ results in spatial distributions that are visually identical to those displayed in Fig.~\ref{fig:com_plot_G}.

\subsubsection{Arrival time}

\begin{figure}[t!]
    \centering
    \includegraphics[width=0.49\linewidth]{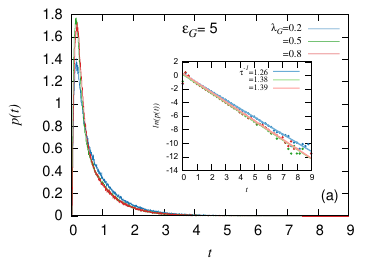} 
    \includegraphics[width=0.49\linewidth]{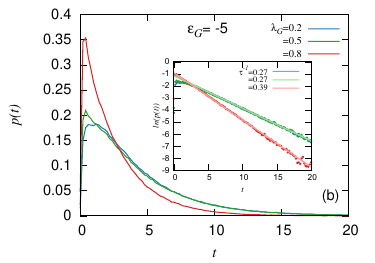}
    \includegraphics[width=0.49\linewidth]{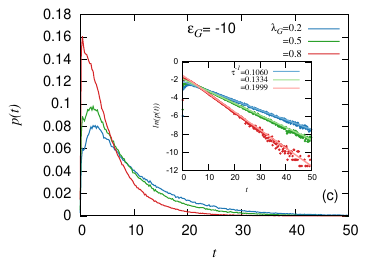}
    \caption{Arrival time distributions for three different Gaussian pair potential~\eqref{eq:gpp} amplitude $\varepsilon_G$ values: (a) $\varepsilon_G=5$, (b) $\varepsilon_G=-5$ and (c) $\varepsilon_G=-10$. In each panel, we compare the distributions for three values of $\lambda_G$, the pair potential range. The insets show semilogarithmic plots of the same distributions as in the main plots. The straight lines are fits to the tails of each distribution; the legends indicate the fitted decay times $\tau$  for each $\lambda_G$.}
    \label{fig:t_e_s}
\end{figure}

We now discuss the arrival time distributions, i.e.\ the time taken by the first particle to reach the destination, for the case where the particles interact via a Gaussian pair potential.
Figure~\ref{fig:t_e_s} shows the probability density of arrival times $p(t)$ for different values of $\varepsilon_G$ and $\lambda_G$.
Note that we do not display results for the noninteracting case where $\varepsilon_G=0$, because these (effectively single-particle) results are already presented in Ref.~\cite{ahmad2025nonequilibrium}.
It is immediately apparent from comparing the ranges of the three different $t$-axes that the typical time for the first particle to travel from $A$ to $E$ increases as we move from repulsive $\varepsilon_G>0$ to attractive $\varepsilon_G<0$. 
This is due to the particles sticking together when the attraction is strong enough.
The resulting pair take a longer time to reach the destination than a single particle on its own. 

The insets of Fig.~\ref{fig:t_e_s} show the temporal dependence of $\ln(p(t))$. Except at very short times, the distributions $p(t)$ have the following exponential form 
\begin{align} 
    p(t)=\chi e^{-\frac{t}{\tau}},
    \label{eq:exp_distrib}
\end{align}  
where the constant $\chi$ (obtained by fitting to the data) is close, but not exactly equal, in value to $1/\tau$.
This is because of the deviations at short times from the exponential distribution.
The exponential form in Eq.~\eqref{eq:exp_distrib} would be exact if the rates $k_{AB}$ and $k_{AD}$ for crossing each of the barriers at $B$ and $D$, respectively, were constant in time.
However, as discussed in Refs.~\cite{fitzgerald2023stochastic, ahmad2025nonequilibrium}, at early times these take different values from their long-time limiting values, so there are deviations from the exponential form in Eq.~\eqref{eq:exp_distrib} at early times $t$.
In Fig.~\ref{fig:t_e_s} we also display exponential fits to the tails of the data. 
We observe that for each $\varepsilon_G$ value, upon increasing $\lambda_G$, the reciprocal of the characteristic time $\tau^{-1}$ increases (see inset legends).

\subsection{Lennard-Jones potential}\label{sec:LJ}

We now discuss results for pairs of Brownian particles interacting via the LJ pair potential in Eq.~\eqref{eq:LJ_GF}.
Unlike for the Gaussian potential, we cannot initialize both particles fully overlapping at the same start position $A$, due to the core repulsion of the LJ potential.
As discussed in Sec.~\ref{sec:IID}, we intialize particle 1 at point $P_1$ and particle 2 at point $P_2$, with coordinates $(x_A+0.5r_{eq}, y_A)$ and $(x_A-0.5r_{eq}, y_A)$, respectively.
Since the LJ equilibrium distance $r_{eq}=2^{\frac{1}{6}}\sigma_{LJ}$ depends on the interaction potential parameter $\sigma_{LJ}$, this raises the natural question of how the dynamics of the system is affected by the choice of initial conditions.
In the following, we vary the value of $\sigma_{LJ}$, with the largest value considered being $\sigma_{LJ}=1$. This sets the maximum distance between the particle positions either side of the initial point $A$ to be $r_{eq}=2^{\frac{1}{6}}\sigma_{LJ}\approx1.12$. Thus, we ensure that the size of the particles is not large enough to cover the entire attractive basin around point $A$ in the external potential.

\begin{figure}[t!]
    \centering
    \includegraphics[width=0.5\linewidth]{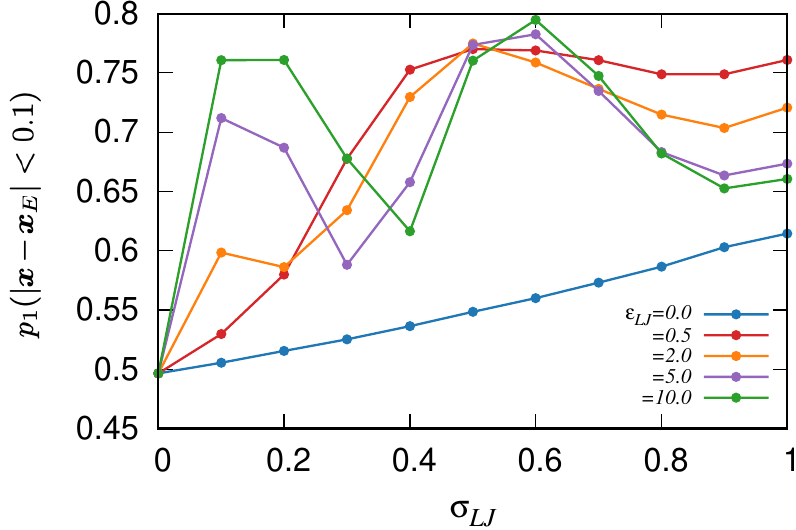}
    \caption{Dependence of the probability $p_1(|\boldsymbol{x}-\boldsymbol{x}_E|<0.1)$ on $\sigma_{LJ}$ for the particle starting at $P_1$ (near $A$) to reach the destination $E$ first, while interacting with the second particle via the LJ potential, for various different $\varepsilon_{LJ}$, as given in the key.}
    \label{fig:p_x1_r}
\end{figure}

To investigate the impact of the initial particle separation $r_{eq}$ on the probability of particle 1 (starting at $P_1$) to reach $E$ first, we calculated this probability for the range of values $\sigma_{LJ}=\{0,0.1,0.2\cdots,1\}$ and for
$\varepsilon_{LJ}=\{0,0.5,2,5,10\}$.
Our results are displayed in Fig.~\ref{fig:p_x1_r}.
When both $\sigma_{LJ}=0$ and $\varepsilon_{LJ}=0$, each particle has the same probability (50\%) of reaching the destination; this case corresponds to two free point-like particles. Upon increasing $\sigma_{LJ}$, the probability of particle 1 (starting at $P_1$, located to the right of the point $A$) to reach the destination increases. 
This is because an increase in the value of $\sigma_{LJ}$ makes the start point $P_1$ slightly closer to the saddle point $D$ that defines the higher energy barrier, and much farther away from the saddle point $B$ that defines the lower energy barrier.
This makes it easier to climb the high energy barrier to reach the destination $E$.
Strikingly, the probability increases further with higher values of $\varepsilon_{LJ}$, but becoming non-monotonic as a function of $\sigma_{LJ}$.
We believe this is due to the fact that for higher values of $\varepsilon_{LJ}$, the pair strongly bind, forming a dimer of length $\sigma_{LJ}$, that is relatively inflexible.
For some dimer lengths, the resulting `rod' is commensurate with the features (minima and valleys) of the external potential, while for other lengths it does not fit well, strongly affecting which way the dimer travels through the potential energy landscape away from the start point.

Because the initial start points $P_1$ and $P_2$ have an impact on the probability of the respective particles making their way to point $E$, we must discuss the location and arrival time statistics for each particle separately.  
Thus, in the next section, we provide a discussion of the location statistics of particles 1 and 2 at the moment if/when the other particle has reached point $E$ first, followed by a discussion of their respective arrival times, for various values of $\varepsilon_{LJ}$ and $\sigma_{LJ}$.

\subsubsection{Location}

\begin{figure*}[t!] 
    \centering
    \includegraphics[width=.98\textwidth]{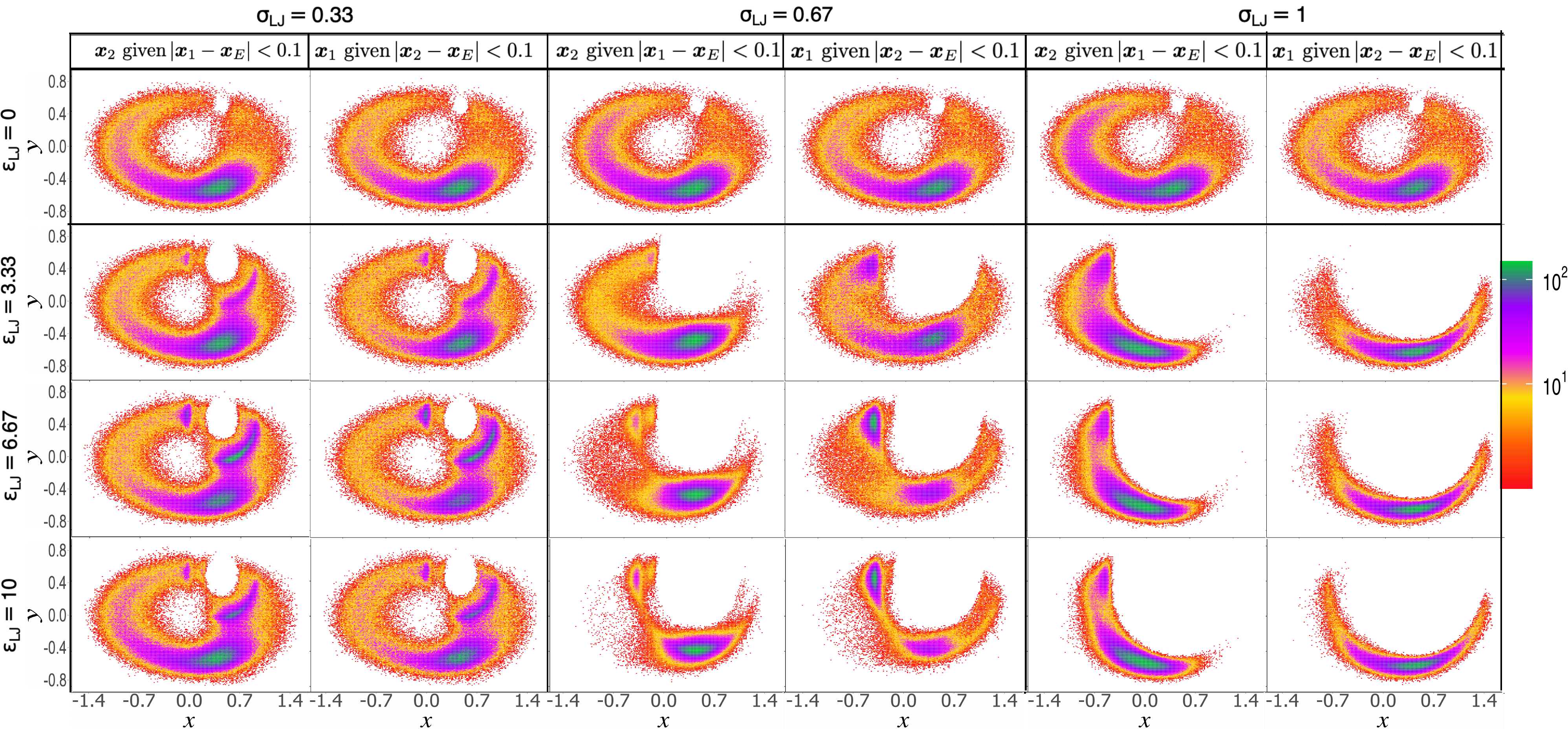}
    \caption{Probability distributions for the position of one particle at the moment when the other has reached the destination point $E$ for the LJ potential in Eq.~\eqref{eq:LJ_GF}, at different values of  $\varepsilon_{LJ}$ (rows)  and  $\sigma_{LJ}$ (columns). Columns 1, 3, and 5 correspond to the distribution of particle 1 (starting at $P_1$) when particle 2 has reached the destination.  Columns 2, 4 and 6 correspond to the distribution of particle 2, when particle 1 has arrived first. These are obtained by averaging over  $5\times10^5$ independent runs.} 
    \label{fig:com_plot_LJ_p1_p2}
\end{figure*}

Figure~\ref{fig:com_plot_LJ_p1_p2} shows the probability distributions for the position of one particle, when the other has reached the destination point $E$. Columns 2, 4, and 6 of Fig.~\ref{fig:com_plot_LJ_p1_p2} refer to the position of particle 1 (starting at $P_1$), when particle 2 has reached $E$ for four values of $\varepsilon_{LJ}$ (rows) and three values of $\sigma_{LJ}$ (columns). Columns 1, 3, and 5 correspond to the position distributions of particle 2, when particle 1 has reached $E$ first, for the corresponding values of $\varepsilon_{LJ}$ and $\sigma_{LJ}$.

In the $\varepsilon_{LJ}=0$ (noninteracting particles) case, the distribution of particle 1, when particle 2 has arrived $E$, and the distribution of particle 2, when particle 1 has reached the point $E$, are very similar for the smaller values of $\sigma_{LJ}$, only becoming noticeably different for the largest value ($\sigma_{LJ}=1$).
This is due to the fact that the start points of the two particles are different, so the distributions are different, even though they are noninteracting.
In the absence of interaction, the distributions of each particle spreads throughout the landscape of $\phi(\boldsymbol{x})$ starting from the local minima at point $A$, making their way to the saddle point $B$; see Refs.~\cite{fitzgerald2023stochastic, ahmad2025nonequilibrium} for further discussion of this effectively single independent dynamics case.
When the LJ diameter is small ($\sigma_{LJ}=0.33$), the distributions 
for the three cases $\varepsilon_{LJ}=\{3.33,6.67,10\}$ do not drastically differ from the non-interacting case $\varepsilon_{LJ}=0$. This is on account of  the particle's diameter being rather small so that the interaction quickly decays to zero. Furthermore, the concentration of the particle that did not reach the destination is mostly concentrated around the local minimum point $A$ and above the saddle point $D$.

For $\sigma_{LJ}=0.67$, we observe that the presence of a non-zero interaction causes a drastic drop in particle concentration around the saddle point $D$ and the appearance of local maximum in probability density in the potential valley between saddle point $B$ and the global minimum point $E$. An increase in the value of $\varepsilon_{LJ}$ makes this structural change in the probability density more dramatic.

For $\sigma_{LJ}=1$ (and nonzero $\varepsilon_{LJ}$), we observe that the probability density concentrates mostly  around the local minimum $A$ in a crescent shape skewed towards the saddle point $B$.

\begin{figure}[t!] 
    \centering
    \includegraphics[width=.5\linewidth]{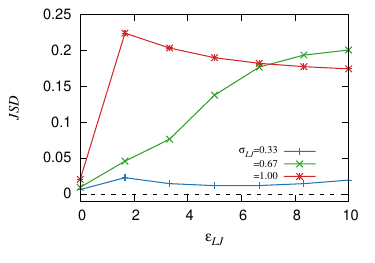}
    \caption{Jensen--Shannon divergence between the probability distributions of the positions of particle 1 when particle 2 has reached the destination $E$ first and the distributions of particle 2, when 1 has arrived at $E$ first, for the LJ potential. A lower JSD indicates greater similarity between the distributions.}. 
    \label{fig:JSD}
\end{figure}

Figure~\ref{fig:JSD} shows the JSD metric \eqref{eq:JSD} for the difference between the probability distributions shown in Fig.~\ref{fig:com_plot_LJ_p1_p2} of the positions of the particles. It is clear that for $\sigma_{LJ}=0.33$ the probability distributions of the two particles are rather similar, reflected by the fact that the value of JSD is relatively small.
In contrast, for a moderate value $\sigma_{LJ}=0.67$, we observe that upon increasing $\varepsilon_{LJ}$ the JSD grows, reflecting the fact that the probability distributions grow increasingly different.
However, for the largest value of $\sigma_{LJ}$ explored here, we observe a maximum in JSD at $\varepsilon_{LJ}=1.67$ followed by a slowly  decreasing trend, suggesting a maximum dissimilarity in the probability distribution of two particles as $\varepsilon_{LJ}$ and $\sigma_{LJ}$ are varied.
These observations made through the lens of the JSD in Fig.~\ref{fig:JSD}, supplement our finding from Fig.~\ref{fig:com_plot_LJ_p1_p2}. 
A particularly striking aspect of Fig.~\ref{fig:JSD} is that the JSD is a non-monotonic function of $\sigma_{LJ}$ for $\varepsilon_{LJ}>7$, highlighting again the non-trivial dynamics of dimers in the potential $\phi(\boldsymbol{x})$, as a function of dimer length $r_{eq}$.

\subsubsection{Arrival time}

\begin{figure}[t!]
    \centering    
    \includegraphics[width=0.5\linewidth]{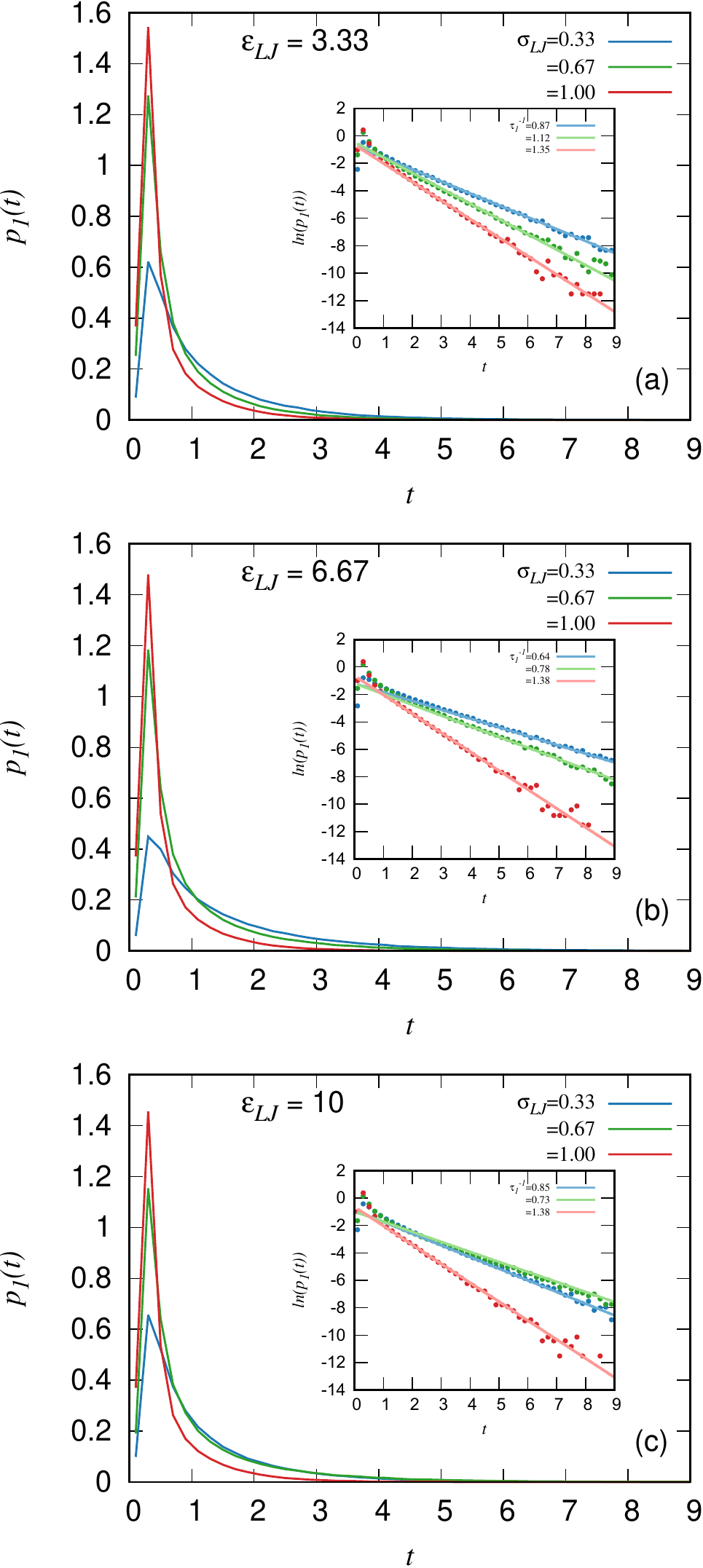}
    \caption{Arrival time distributions $p_1(t)$ of the particle 1 (starting at $P_1$, to the right of point $A$), subject to the LJ potential for (a) $\varepsilon_{LJ}=3.33$, (b) $\varepsilon_{LJ}=6.67$ and (c) $\varepsilon_{LJ}=10.0$, respectively. In each panel,  we compare  the distributions for different values of $\sigma_{LJ}$. The insets show the semilogarithmic plots of the same distributions in the main plots. The straight lines are fits to the tails of each distribution; the legends indicate the fitted decay times $\tau_1$. The straight lines are fits to the tails of each distribution; the legends indicate the fitted decay times $\tau$ for each corresponding $\sigma_{LJ}$.}
    \label{fig:p1_t_e_s}
\end{figure}

\begin{figure}[t!]
    \centering  
    \includegraphics[width=0.5\linewidth]{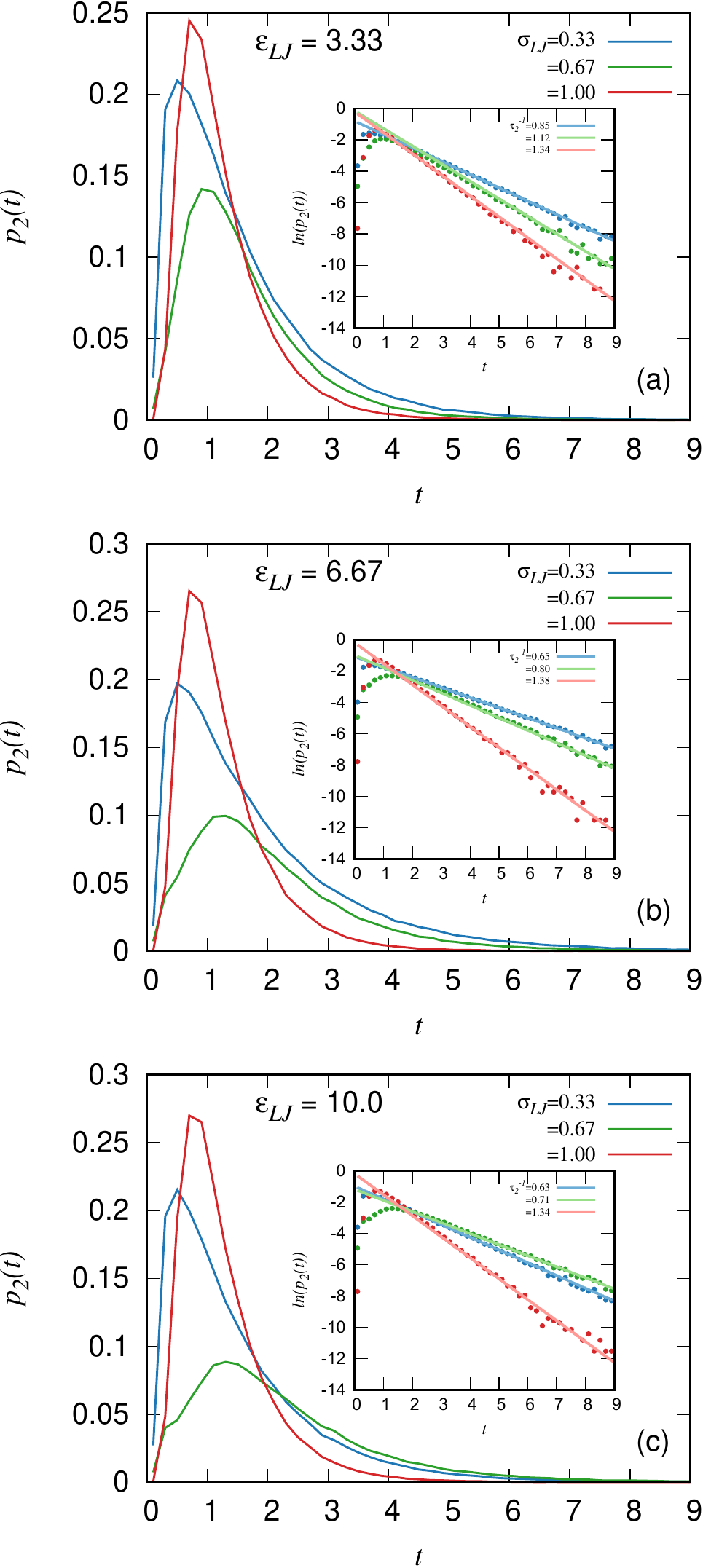}
    \caption{Arrival time distributions $p_2(t)$ of the particle 2 (starting at $P_2$, to the left of point $A$), with everything else the same as in Fig.~\ref{fig:p1_t_e_s}.}
    \label{fig:p2_t_e_s}
\end{figure}

Next, we examine the temporal evolution of our two-particle system, illustrating how LJ interactions induce changes in the statistics of the arrival times.  Figure~\ref{fig:p1_t_e_s} shows the arrival time distribution $p_1(t)$ of particle 1, when particle 2 has reached $E$ first, whereas Fig.~\ref{fig:p2_t_e_s} shows the arrival time distribution $p_2(t)$ of particle 2, when particle 1 has reached $E$ first.

From Fig.~\ref{fig:p1_t_e_s}, we observe that the arrival time distribution $p_1(t)$ is strongly peaked around small times for all $\varepsilon_{LJ}$. However, this peak increases in magnitude upon increasing $\sigma_{LJ}$.
The strong peak at relatively small times is to be expected, since particle 1 starts slightly closer to the destination $E$, allowing it to reach it earlier. 

The insets of Fig.~\ref{fig:p1_t_e_s} show semilogarithmic plots of $p_1(t)$ for the same data in the primary plots. Except at very short times, the distributions are well fitted by straight lines, suggesting that the tails of $p_1(t)$ can be modeled by exponential distributions, similar to Eq.~\eqref{eq:exp_distrib}
\begin{align}
    p_1(t)=\chi_{1} e^{-\frac{t}{\tau_1}},
    \label{eq:exp1_distrib}
\end{align}  
where the constant $\chi_1$ is not exactly equal to $1/\tau_1$ but is~$\approx1/\tau_1$, due to deviations at short times. These exponential fits are shown as straight lines in the Fig.~\ref{fig:p1_t_e_s} insets.

Given the behavior of the spatial distributions in Fig.~\ref{fig:com_plot_LJ_p1_p2}, one might expect $p_1(t)$ to exhibit a strong dependence on  $\sigma_{LJ}$, but infact this is not the case.
An increase in $\sigma_{LJ}$ for any given $\varepsilon_{LJ}$ does give an increase in $\tau_1^{-1}$, as shown in the insets, but the variations between the various panels (corresponding to changing the value of $\varepsilon_{LJ}$) leads only to small variations.

Figure~\ref{fig:p2_t_e_s} shows that the arrival time distribution $p_2(t)$ is broader than the distribution of $p_1(t)$ for all $\varepsilon_{LJ}$ explored here. Furthermore, also in this case there is a dependence of the $p_2(t)$ distribution on $\sigma_{LJ}$. 
In the insets of Fig.~\ref{fig:p2_t_e_s}, we show the semilogarithmic plot of  $p_2(t)$, for the same data set utilized in the primary plots. These plots again appear to be straight lines, except at very short times, of similar nature of exponential distribution defined in Eq.~\eqref{eq:exp_distrib} and hence the updated distribution $p_2(t)$ for particle 2 reads
\begin{align}
    p_2(t)=\chi_2 e^{-\frac{t}{\tau_2}},
    \label{eq:exp2_distrib}
\end{align}  
where the constant $\chi_2$,  $\tau_2$ are defined in the same way.

Comparing the arrival time distributions $p_1(t)$ and $p_2(t)$, for particles 1 and 2 respectively, displayed in Fig.~\ref{fig:p1_t_e_s} and Fig.~\ref{fig:p2_t_e_s}, indicates that particle 1 reaches the destination more quickly than particle 2, which is to be expected, given the starting position $P_1$ is closer to the destination point $E$.

\begin{figure}[t!]
    \centering  
    \includegraphics[width=0.5\linewidth]{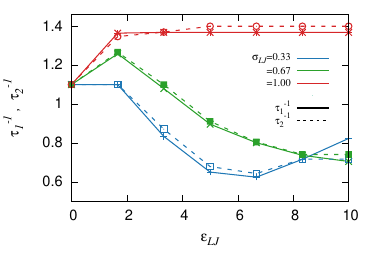}
    \caption{Dependence of the inverse relaxation times  $\tau_1^{-1}$ and $\tau_2^{-1}$ on the interaction strength $\varepsilon_{LJ}$ of LJ-pair potential. The different colors correspond to different $\sigma_{LJ}$.}
    \label{fig:tau12_e_s}
\end{figure}

Figure~\ref{fig:tau12_e_s} shows the dependence of the time relaxation coefficients $\tau_1$ and $\tau_2$ defined in Eqs.~\eqref{eq:exp1_distrib} and \eqref{eq:exp2_distrib} on the interaction parameter $\varepsilon_{LJ}$ of the LJ-pair potential. When $\sigma_{LJ}$ is small compared to the size of the wells in the external potential $\phi(\boldsymbol{x})$ and the distance between the wells, we see from Fig.~~\ref{fig:tau12_e_s} that particle interactions increase the overall time for particles to reach the destination $E$. However, there is a nonmonotonic effect, and they are the slowest when $\varepsilon_{LJ}\approx6$. 

When $\sigma_{LJ}=1$, the values of both $\tau$ remain largely unaffected by variations in $\varepsilon_{LJ}$, that is, as $\sigma_{LJ} \to |\bm{x}_C-\bm{x}_B|$, the values of $\tau_k$ are relatively insensitive to changes in $\varepsilon_{LJ}$, potentially implying underlying stability or saturation in the system response.
In fact, when $\sigma_{LJ}$ is large, particle interactions speed up the particles to reach the destination.

\section{Concluding remarks} \label{sec:6}

We have studied the (noise) activated barriers-crossing process, and specifically the influence of particle interactions, of a pair of particles crossing barriers in systems with competing pathways.
The external potential $\phi(\boldsymbol{x})$ considered is of interest because standard reaction rate theory (RRT) predicts that the most probable route is to cross the lowest-energy barrier \cite{fitzgerald2023stochastic, ahmad2025nonequilibrium}.
Thus, the expectation from RRT is that this longer route should dominate the dynamics of the system.
However, even when there is a single particle, the standard RRT framework can fail to accurately capture the dynamics, becoming critically dependent on the timescale over which the system is observed or sampled \cite{fitzgerald2023stochastic, ahmad2025nonequilibrium}.
Such barrier-crossing problems involving competing pathways have broad implications.

Here, we have considered the case when there are a pair of particles available to cross the barriers and focussed on the influence of particle interactions on the (rare event) barrier-crossing processes.
We have explored how interactions affect the choice of route taken and the time taken.
Two well-known interaction potentials, the Gaussian pair potential and the LJ potential, were considered.
In the LJ case, the interacting particles cannot overlap, so cannot start at exactly the same point, as was considered in the Gaussian pair potential case.
Taken all together, our results show that particle interactions can play a crucial role in determining which particle crosses one of the barriers first and how long it takes.
When there are strong attractive interactions, the pair of particles bind together and can sometimes be treated as a single larger particle.
However, the presence of a repulsive core to the interaction, which introduces an additional length-scale,  can have a major additional influence.
When such particles bind together, the length of the resulting dimer and how that fits (or not) within the potential wells of the confining potential has a strong impact on the resulting first-passage time statistics.
When interactions are purely repulsive,  we generally find barrier-crossing occurs faster, since one particle helps to `push' the other over the barrier.

The main findings of our results on barrier-crossing of pairs of particles are that: (i) the initial conditions of the particles in the external potential matter. (ii) Particle interactions introduce additional physics, which can greatly change the barrier crossing rates and choice of routes. (iii) Particle interactions not only introduce additional (binding) energy considerations, they also introduce additional length-scales, resulting in a complex interplay with the geometry of the confining potential.
We expect these observations to have important implications for our understanding of chemical and biological reactions, for example stochastic target-search and confined-transport processes in which several particles simultaneously explore a structured energy landscape \cite{dror2011pathway,tiwary2015kinetics}. 
Competing pathways and transition times are important in molecular recognition and drug–target binding, where molecular crowding can modify recognition kinetics \cite{ellis2001macromolecular}.

\begin{acknowledgments}
GA gratefully acknowledges Loughborough University Doctoral College UK and HEC Pakistan for funding support under ``Overseas Scholarships for PhD in Selected Fields Phase-III Batch 3''.
\end{acknowledgments}


%

\end{document}